\documentclass[aps, prstab, reprint, showpacs, amsmath, amssymb]{revtex4-1}

\usepackage{amsmath,amscd}
\usepackage{bm}
\usepackage[english]{babel}
\usepackage{graphicx}
\usepackage{soul}
\usepackage[usenames]{color}
\usepackage{physics}
\DeclareMathOperator{\w}{\omega}
\DeclareMathOperator{\ve}{\varepsilon}
\DeclareMathOperator{\R}{\rho}

\renewcommand{\Im}{\mathop{\mathrm{Im}}\nolimits}

\begin{document}
\title{ Diffraction of a Waveguide Mode at the Transverse Boundary of Magnetized Plasma }

\author{Sergey N. Galyamin}
\email{s.galyamin@spbu.ru}
\affiliation{Saint Petersburg State University, 7/9 Universitetskaya nab., St. Petersburg, 199034 Russia}

\date{\today}

\begin{abstract}
Here we develop a general theory of mode transformation (diffraction) at the flat transverse boundary between cold magnetized electron plasma and isotropic vacuum-like medium inside a circular waveguide.
The obtained results can be also directly applied to the narrow-band Cherenkov radiation generated in plasma (or in isotropic medium) by a moving charged particle bunch.
\end{abstract}

\maketitle

\section{Introduction}

In last decades, interaction of charged particle bunches with plasma guiding structures has attracted an essential attention.
For example, electron bunch injected in an open plasma waveguide (at certain additional conditions usually referred to as ``density duct'') can effectively generate electromagnetic (EM) waves of whistler range (due to Cherenkov radiation mechanism), the latter can be in turn effectively guided by the duct~\cite{Zaboronkova2002,Zaboronkova2007}.
Whistler waves themselves are considered as a promising tool for wave diagnostics of near cosmic space and manipulation of properties of EM radiators (antennas) in magnetized plasma. 
It should be also noted that problems of guided VLF propagation in a waveguide earth - ionosphere were extensively investigated much earlier~\cite{Buddenb, MNRb}, where appropriate methods of analysis have been developed. 
Moreover, in recent years an essential progress has been achieved in particle acceleration within Plasma Wakefield Acceleration (PWFA) scheme~\cite{Blum07}.
One of the most promising configuration of plasma -- hollow plasma channel -- has been recently investigated in a series of papers.
For example, in~\cite{Gessner16} both electron and positron acceleration with gradients that are orders of magnitude larger than those achieved in conventional accelerators has been shown.
High-gradient plasma wakefields can be also used for development of plasma-based radiation sources.
If magnetized plasma is used as a radiator, the external magnetic field can serve to additionally affect the properties of generated fields.  
Moreover, in this case radiation should be extracted from the radiator through some interface, therefore certain boundary problem occurs.
Boundary problems with various complicated dielectric media in a circular waveguide resulting in a series of prospective effects (for example, enhancement of radiation intensity, self-acceleration of the bunch, reversed Cherenkov-transition radiation) were also actively investigated~\cite{AT11, AT12, AT13, AT14, AT19, AT19RPC}.
In particular, paper~\cite{AT19RPC} dealt with a limiting case of a strongly magnetized plasma (the external magnetic field tends to infinity).
However, it is of interest to consider the general case of arbitrary external magnetic field where plasma possesses both anisotropy (similar to the case of strong magnetization) and gyrotropy.    

In this paper we consider a problem where a longitudinally magnetized plasma fills a half of regular cylindrical waveguide while the second half is vacuum (or vacuum-like isotropic medium).
If a moving particle bunch excites this plasma, Cherenkov radiation in the form of a discrete set of waveguide modes, each having certain Cherenkov frequency, is generated.
In some cases, a single Cherenkov frequency regime can be organized.
Therefore, for the sake of shortness and simplicity of derivations, we consider the case where single waveguide mode incidents the interface between the plasma and vacuum.

\section{Problem formulation}

%
\begin{figure}[b]
\centering
\includegraphics[width=0.4\textwidth]{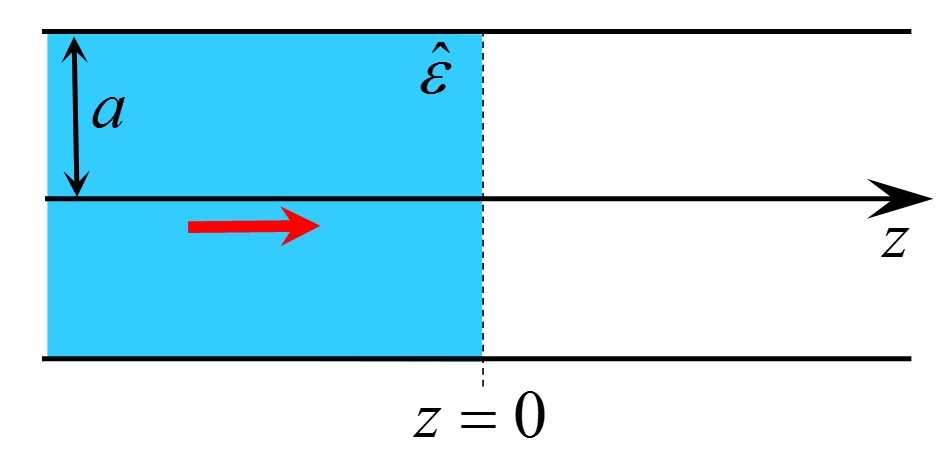}
\caption{\label{fig:geom} Geometry of the problem and main notations.}
\end{figure}
%

We consider a circular waveguide of radius $a$ a half of which is filled with cold magnetized plasma, see Fig.~\ref{fig:geom}.
Plasma is described by the following tensor of dielectric permittivity:
\begin{equation}
\label{eq:tensor}
\hat{\ve}
=
\left(
\begin{array}{ccc}
{\ve_{\bot}}&{-ig}&0 \\
{ig}&{\ve_{\bot}}&0 \\
0&0&{\ve_{\parallel}}
\end{array}
\right)
\end{equation}
and magnetic permeability
$\mu{=}1$.
The
$g$
component causes gyrotropy, while the inequality of components
$\ve_{\bot}$
and 
$\ve_{\parallel}$
creates uniaxial anisotropy.
Here,
$\ve_{\bot}$,
$g$
and
$\ve_{\parallel}$
are assumed to be frequency dependent, i.e., frequency dispersion is taken into account.
A model for a cold electron plasma in an external magnetic field is utilized here.
The frequency dependence of the components of the permittivity tensor in such a medium is described by the following expressions
\cite{Gb, GKT13}:
\begin{equation}
\begin{aligned}
&\ve_{\bot}(\w)
=
1 - \frac{\w_p^2 (\w + i\nu)}{\w \left[ (\w + i\nu)^2 - \w_h^2 \right]}, \\
&g(\w)
=
\frac{- \w_p^2{\w_h}}{\w \left[ (\w + i\nu)^2 - \w_h^2 \right]}, \\
&\ve_{\parallel}(\w)
= 1 - \frac{\w_p^2}{\w^2 + i\w\nu},
\end{aligned}
\label{eq:dispersion}
\end{equation}
where
$\w_p^2{=}4\pi Ne^2{/}m$
is the plasma frequency 
($N$
is the electron density, and
$e$
and
$m$
are the electron charge and the electron mass, respectively),
$\w_h{=}|e|H_{ext}{/}(mc)$
is a ``gyrofrequency'' 
($H_{ext}$
is the external magnetic field directed along $z$-axis) and
$\nu$
is the effective collision frequency.
The second half of the waveguide is vacuum while the outer walls are perfectly conductive.

\subsection{ Modes of plasma filled waveguide }

Normal waves (modes) of a regular plasma filled waveguide are determined as follows.
The problem is solved in the frequency domain so that Fourier integral decomposition is used, for example:
\begin{equation}
H_{\varphi}
=
\int\nolimits_{-\infty}^{+\infty } H_{\w \varphi} e^{ -i \omega t }\,d\omega.
\end{equation}
Maxwell equations without external sources in infinite plasma medium~\eqref{eq:tensor} 
\begin{equation}
\label{Maxwell}
\begin{aligned}
\mathrm{rot}\vec{E}_{\w} &= i k_0 \vec{E}_{\w}, &
\quad
\mathrm{rot}\vec{H}_{\w} &= - i k_0 \vec{D}_{\w}, \\
\mathrm{div}\vec{H}_{\w} &= \mathrm{div}\vec{B}_{\w} = 0, &
\quad
\mathrm{div}\vec{D}_{\w} &= 0,
\end{aligned}
\end{equation}
with material relations 
\begin{equation}
\label{eq:matrel}
\begin{aligned}
\vec{D}_{\w}&=\hat{\ve}\vec{E}_{\w}= \vec{e}_{z} \ve_{\parallel} E_{\w z} +\\ 
&+ \vec{e}_{r}[\ve_{\bot} E_{\w r} - i g E_{\w \varphi} ] + 
\vec{e}_{\varphi}[\ve_{\bot} E_{\w \varphi} + i g E_{\w r} ],
\end{aligned}
\end{equation}
can be handled as follows.
Let us introduce the following vector of unknowns:
\begin{equation}
\label{vector}
\vec{\mathcal{E}} =
\left(
\begin{array}{c}
H_{\w \varphi} \\
E_{\w r} \\
E_{\w \varphi} \\
H_{\w r}
\end{array}
\right),
\end{equation}
while the longitudinal components are calculated as follows:
\begin{equation}
\label{eq:components}
\begin{aligned}
H_{\w z} &= \frac{ 1 }{ i k_0 r } \pdv{ \left(r E_{\w \varphi }\right) }{ r }, & 
E_{\w z} &= \frac{ i }{ k_0 \ve_{\parallel} r } \pdv{ \left( r H_{\w \varphi} \right) }{ r }. 
\end{aligned}
\end{equation}
After a series of simple but bulky transformation we arrive from~\eqref{Maxwell} to the following equation for~\eqref{vector}:
\begin{equation}
\label{opeq}
\pdv{\vec{\mathcal{E}}}{z} - i \hat{\mathcal{L}}_r \vec{\mathcal{E}} = 0,
\end{equation}
where $\hat{\mathcal{L}}_r$ is the transverse operator of the problem:
\begin{widetext}
\begin{equation}
\label{op}
\hat{\mathcal{L}}_r = 
\left(
\begin{array}{cccc}
0 & k_0 \ve_{\bot} & -i k_0 g & 0 \\
k_0+\frac{1}{k_0\ve_{\parallel}}\Delta_r & 0 & 0 & 0 \\
0 & 0 & 0 & -k_0 \\
0 & -i k_0 g & -k_0 \ve_{\bot} - \frac{1}{k_0} \Delta_r & 0 \\
\end{array}
\right),
\end{equation}
\end{widetext}
where
\begin{equation}
\Delta_r \equiv \pdv{\left( \frac{1}{r}\pdv{(r\cdot)}{r}\right) }{r}.
\end{equation}
Boundary conditions at the waveguide walls (PEC) are:
\begin{equation}
\label{bc}
E_{\w \varphi} \eval_{r=a} = 0,
\quad
\pdv{H_{\w \varphi}}{r} + \frac{H_{\w \varphi}}{r} \eval_{r=a} = 0.
\end{equation}

Let us consider an eigenvalue problem for $\hat{\mathcal{L}}_r $:
\begin{equation}
\label{Shturm}
\hat{\mathcal{L}}_r \vec{U} = \lambda \vec{U},
\quad
\vec{U} = 
\left(
\begin{array}{c}
u_1 \\
u_2 \\
u_3 \\
u_4 \\
\end{array}
\right),
\end{equation}
where $\lambda$ is an eigenvalue, $\vec{U}$ is an eigenvector and the following boundary conditions should be fulfilled (a consequence of~\eqref{bc}):
\begin{equation}
\label{bcu}
\pdv{u_1}{r} + \frac{u_1}{r} \eval_{r=a} = 0,
\quad
u_3 \eval_{r=a} = 0.
\end{equation}
After transformations the following coupled system occurs:
\begin{equation}
\label{coupled}
\left\{
\begin{aligned}
&\left(
\Delta_r + k_0^2 \frac{\ve_{\bot}^2 - g^2}{\ve_{\bot}} - \lambda^2 
\right)
u_3  + i \lambda k_0 \frac{ g }{ \ve_{\bot} } u_1 = 0, \\
&\left(
\Delta_r + k_0^2 \ve_{\parallel} - \frac{\ve_{\parallel}}{\ve_{\bot}}\lambda^2 
\right)
u_1 - i \lambda k_0 \ve_{ \parallel } \frac{ g }{ \ve_{\bot} } u_3 = 0, 
\end{aligned}
\right.
\end{equation}
and
\begin{equation}
\label{u2u4}
u_2 = \frac{\lambda}{k_0\ve_{\bot}}u_1 + \frac{ig}{\ve_{\bot}}u_3,
\quad
u_4 = -\frac{\lambda}{k_0}u_3.
\end{equation}
System~\eqref{coupled} can be solved as follows.
Let us rewrite~\eqref{coupled}:
\begin{equation}
\label{eqGamma}
\hat{a}\left(\begin{array}{c} u_1 \\ u_3 \end{array}\right)
+
\hat{\Gamma}\left(\begin{array}{c} u_1 \\ u_3 \end{array}\right) = 0,
\end{equation}
where $\hat{a} = \mathrm{diag}\left(\Delta_r, \Delta_r \right)$,
\begin{equation}
\hat{\Gamma} = \left(
\begin{array}{cc} k_0^2 \ve_{\parallel} - \frac{\ve_{\parallel}}{\ve_{\bot}}\lambda^2 &
- i \lambda k_0 \ve_{ \parallel } \frac{ g }{ \ve_{\bot} } \\ 
i \lambda k_0 \frac{ g }{ \ve_{\bot} } & 
k_0^2 \frac{\ve_{\bot}^2 - g^2}{\ve_{\bot}} - \lambda^2 
\end{array}
\right).
\end{equation}
Matrix $\hat{\Gamma}$ can be diagonalized in a standard way: an eigenvalues $\xi^2$ and eigenvectors $\vec{w}$ of the following problem should be found:
\begin{equation}
\hat{\Gamma}\vec{w} = \xi^2\vec{w},
\quad
\vec{w} = \left(\begin{array}{c} u_1\\u_3\end{array}\right).
\end{equation}
One obtains the following equation for $\xi^2$:
\begin{equation}
\label{oewaves}
\begin{aligned}
&\left(k_0^2 \ve_{\parallel} - \frac{\ve_{\parallel}}{\ve_{\bot}}\lambda^2 - \xi^2 \right)
\left(k_0^2 \frac{\ve_{\bot}^2 - g^2}{\ve_{\bot}} - \lambda^2 - \xi^2 \right) - \\
&-
\lambda^2 k_0^2 \ve_{\parallel} \left( \frac{\ve_{\parallel}}{\ve_{\bot}} \right)^2 = 0,
\end{aligned}
\end{equation}
which is known equation for transverse wavenumbers of ordinary (``$o$'') and extraordinary (``$e$'') plasma waves~\cite{GKT13}, while $\lambda$ plays a role of a longitudinal wavenumber (this will be clarified below), therefore $\xi^2 = s_{o,e}^2$ where:
\begin{equation}
\label{soe}
\begin{aligned}
&s_{o,e}^{2}
{=}
\frac{ 1 }{ 2 \ve_{\bot} }
\left[
\vphantom{
\sqrt{ \left[ ( \ve_{\bot}^2 - g^2 - \ve_{ \bot } \ve_{ \parallel } ) \beta^2 - \ve_{ \bot } + \ve_{ \parallel } \right]^2
+
4 \beta^2 g^2 \ve_{ \parallel } }
}
k_0^2( \ve_{\bot}^2 - g^2 + \ve_{ \bot } \ve_{ \parallel } ) - ( \ve_{ \bot } + \ve_{ \parallel } )\lambda^2 \right.
\pm \\
&\left.
\sqrt{ \left[ k_0^2 ( \ve_{\bot}^2 - g^2 - \ve_{ \bot } \ve_{ \parallel } ) - ( \ve_{ \bot } - \ve_{ \parallel } ) \lambda^2 \right]^2
+
4  k_0^2 \lambda^2 g^2 \ve_{ \parallel } }
\right],
\end{aligned}
\end{equation}
with ``--'' corresponding to ``$o$'' and ``+'' corresponding to ``$e$''.
Eigenvectors $\vec{w}^o$ and $\vec{w}^e$ form a matrix $\hat{T}$
\begin{equation}
\label{T}
\begin{aligned}
\hat{T} &= 
\left( \begin{array}{cc} \vec{w}^o & \vec{w}^e \end{array} \right)= \\
&=
\left( \begin{array}{cc} \frac{ i \lambda k_0 \ve_{\parallel} \frac{g}{\ve_{\bot}}}{k_0^2 \ve_{\parallel} - \frac{\ve_{\parallel}}{\ve_{\bot}} \lambda^2 - s_o^2} & 1 \\
1 & \frac{ - i \lambda k_0 \frac{g}{\ve_{\bot}}}{k_0^2 \frac{\ve_{\bot}^2 - g^2}{\ve_{\bot}} - \lambda^2 - s_e^2} \end{array} \right),
\end{aligned}
\end{equation}
which diagonalizes the matrix $\hat{\Gamma}$,
\begin{equation}
\hat{D} = \hat{T}^{-1}\hat{\Gamma}\hat{T} = \mathrm{diag}\left( s_o^2, s_e^2 \right),
\end{equation} 
and equation~\eqref{eqGamma} (after left multiplication by $\hat{T}^{-1}$) takes the form:
\begin{equation}
\label{eqD}
\hat{a}\vec{\w} + \hat{D}\vec{\w} = 0,
\quad
\vec{\w} = \hat{T}^{-1}\vec{w}.
\end{equation}
Solution of~\eqref{eqD} in the region $0<r<a$ is found straightforwardly,
\begin{equation}
\label{eqDsol}
\vec{\w} = \left(\begin{array}{c} C^o J_1(rs_o) \\ C^e J_1(rs_e) \end{array} \right),
\end{equation} 
where $C^o$ and $C^e$ are some constants, and we obtain the solution of the system~\eqref{coupled}:
\begin{widetext}
\begin{equation}
\label{wsol}
\vec{w} = \left(\begin{array}{c} u_1 \\ u_3 \end{array} \right) =
\hat{T}\vec{\w} = 
\left(
\begin{array}{c}
C^o 
\frac{ i \lambda k_0 \ve_{\parallel} \frac{g}{\ve_{\bot}}}
{k_0^2 \ve_{\parallel} - \frac{\ve_{\parallel}}{\ve_{\bot}}\lambda^2 - s_o^2} 
J_1(rs_o) + C^e J_1(rs_e)\\
C^o J_1(rs_o) 
- 
C^e 
\frac{ i \lambda k_0 \frac{g}{\ve_{\bot}}}{k_0^2 \frac{\ve_{\bot}^2 - g^2}{\ve_{\bot}} - \lambda^2 - s_e^2}
J_1(rs_e) \\
\end{array}
\right).
\end{equation}
\end{widetext}
Applying~\eqref{bcu} to~\eqref{wsol} we arrive at homogeneous linear system of two equations for $C^{o,e}$, the determinant of this system should equal to zero, and we obtain the following dispersion equation for $\lambda$:
\begin{equation}
\label{lambdasol}
\begin{aligned}
&s_oJ_0(as_o)J_1(as_e)
\frac{k_0^2 \frac{\ve_{\bot}^2 - g^2}{\ve_{\bot}} - \lambda^2 - s_o^2}
{k_0^2\frac{\ve_{\bot}^2 - g^2}{\ve_{\bot}} - \lambda^2 - s_e^2} - \\
&- s_eJ_0(as_e)J_1(as_o) = 0
\end{aligned}
\end{equation}
with solutions called $\lambda_p$, $p = \pm 1, \pm 2, \ldots$.
Each of~\eqref{bcu} allows us to express $C^o$ through $C^e$; for example, it is convenient to introduce a ``polarization coefficient'' $\R_p$:
\begin{equation}
\label{rho}
\R_p = \frac{C^o_p}{C^e_p} = 
\frac{J_1(as_{ep})}{J_1(as_{op})}\frac{i \lambda_p k_0 \frac{g}{\ve_{\bot}}}{k_0^2 \frac{\ve_{\bot}^2 - g^2}{\ve_{\bot}} - \lambda_p^2 - s_{ep}^2}.
\end{equation}

Introducing the following designations
\begin{equation}
\label{Qoe}
\begin{aligned}
Q_{op} &= \frac{ i \lambda_p k_0 \ve_{\parallel} \frac{g}{\ve_{\bot}}}{k_0^2 \ve_{\parallel} - \frac{\ve_{\parallel}}{\ve_{\bot}}\lambda_p^2 - s_{op}^2}, \\
Q_{ep} &= \frac{ - i \lambda_p k_0 \frac{g}{\ve_{\bot}}}{k_0^2 \frac{\ve_{\bot}^2 - g^2}{\ve_{\bot}} - \lambda_p^2 - s_{ep}^2},
\end{aligned}
\end{equation}
one can finally write down the solution for eigenvector $\vec{U}_p$ of the transverse operator~\eqref{op}:
\begin{widetext}
\begin{equation}
\label{Usol}
\vec{U}_p = 
\left(
\begin{array}{c}
u_{1p} \\
u_{2p} \\
u_{3p} \\
u_{4p} \\
\end{array}
\right)
=
\left(
\begin{array}{c}
\R_p Q_{op} J_1(rs_{op}) + J_1(rs_{ep}) \\
\R_p \left[\frac{\lambda_p Q_{op} }{k_0 \ve_{\bot}} +\frac{ig}{\ve_{\bot}} \right] J_1(rs_{op}) 
+ \left[\frac{\lambda_p }{k_0 \ve_{\bot}} +\frac{igQ_{ep} }{\ve_{\bot}} \right] J_1(rs_{ep}) \\
\R_p J_1(rs_{op}) + Q_{ep} J_1(rs_{ep}) \\
-\frac{\lambda_p}{k_0}\left[ \R_p J_1(rs_{op}) + Q_{ep} J_1(rs_{ep}) \right] \\
\end{array}
\right).
\end{equation}
\end{widetext}

Solution for $\vec{\mathcal{E}}$ can be decomposed over eigenvectors $\vec{U}_p$:
\begin{equation}
\label{decomp}
\vec{\mathcal{E}} = 
\sum\limits_{p=-\infty}^{+\infty}
A_p(z)\vec{U}_p(r),
\end{equation}
where $A_p(z)$ unknown functions.
Substituting~\eqref{decomp} to~\eqref{opeq} and utilizing the formula $\hat{\mathcal{L}}_r\vec{U}_p = \lambda_p\vec{U}_p$ one obtains:
\begin{equation}
\label{Asol}
A_p(z) = a_p \exp \left[ i \lambda_p z \right],
\end{equation}
where $a_p$ some constants.
It is clear that physical sense of $\lambda_p$ is a longitudinal wavenumber.
It is expedient to recall $A_p\vec{U}_p = a_p\vec{U}_p\exp\left[ i \lambda_p z \right]$ the normal wave or the mode of the considered plasma waveguide and $a_p$ the magnitude of the mode. 
In the lossless case ($\nu=0$) modes with $\Im{\lambda_p} = 0$ are propagating, otherwise they are evanescent.
Propagating modes with $\lambda_p > 0 $ propagate in positive $z$-direction, while modes with $\lambda_p < 0 $ -- in negative $z$-direction.
We will suppose that for positive $p$: $ \lambda_p > 0$ (if real) or $\Im\lambda_p >0$ (if imaginary).  

It is also useful to mention the following properties (when $p$ changes to $-p$):
\begin{equation}
\label{p2-p}
\begin{aligned}
\R_{-p} = - \R_p, \,\,
Q_{o-p} = - Q_{op}, \,\, 
Q_{e-p} = - Q_{ep},  \\
u_{1-p} = u_{1p}, \,\,
u_{2-p} = - u_{2p}, \,\,
u_{3-p} = - u_{3p}, \\
u_{4-p} = u_{4p}, \,\,
N_{-p} = - N_p.
\end{aligned}
\end{equation}

\subsection{ Orthogonality of modes of plasma filled waveguide }

As it is known~\cite{Mittrab}, in the considered magnetized plasma modes possess biorthogonality.
To obtain this property explicitly, one should construct eigenvectors of the conjugated operator~$\hat{\mathcal{L}}_r^*$.
Scalar product $\langle \cdot, \cdot \rangle$ is defined as follows:
\begin{equation}
\label{scalprod}
\langle \vec{F}_1, \vec{F}_2 \rangle
=
\int\limits_0^a \left( \vec{F}_1, \vec{F}_2 \right) r\,dr,
\end{equation}
where $\vec{F}_{1,2}$ are some vector-functions from $L_2$ functional space.
By definition,
\begin{equation}
\label{opconjdef}
\langle \hat{\mathcal{L}}_r \vec{U}, \vec{V} \rangle = 
\langle \vec{U}, \hat{\mathcal{L}}_r^* \vec{V} \rangle,
\end{equation}
where $\vec{U}$ is an eigenvector of $\hat{\mathcal{L}}_r$ and $\vec{V}$ is a vector with components from $L_2$ functional space.
After standard derivations we obtain:
\begin{widetext}
\begin{equation}
\label{opconj}
\hat{\mathcal{L}}_r^* = 
\left(
\begin{array}{cccc}
0 & k_0+\frac{1}{k_0\ve_{\parallel}}\Delta_r & 0 & 0 \\
k_0 \ve_{\bot} & 0 & 0 & -i k_0 g \\
-i k_0 g & 0 & 0 & -k_0 \ve_{\bot} - \frac{1}{k_0} \Delta_r \\
0 & 0 &  -k_0 & 0 \\
\end{array}
\right),
\quad
\vec{V} = 
\left(
\begin{array}{c}
v_1 \\
v_2 \\
v_3 \\
v_4 \\
\end{array}
\right)
\end{equation}
\end{widetext}
(that is columns of $\hat{\mathcal{L}}_r^*$ coincide with rows of $\hat{\mathcal{L}}_r$ and vice versa), where the domain of $\hat{\mathcal{L}}_r^*$ should be determined as follows:
\begin{equation}
\label{bcv}
\pdv{v_2}{r} + \frac{v_2}{r} \eval_{r=a} = 0,
\quad
v_4 \eval_{r=a} = 0.
\end{equation}
Vector $\vec{V}$ satisfying~\eqref{bcv} is an eigenvector of $\hat{\mathcal{L}}_r^*$.
An eigenvalue problem
\begin{equation}
\label{Shturmv}
\hat{\mathcal{L}}_r^* \vec{V} = \lambda^* \vec{V},
\end{equation}
has the following solution:
\begin{equation}
\label{Vsol}
\lambda_p^* = \lambda_p,
\quad
\vec{V}_p = 
\left(
\begin{array}{c}
v_{1p} \\
v_{2p} \\
v_{3p} \\
v_{4p} \\
\end{array}
\right)
=
\left(
\begin{array}{c}
u_{2p} \\
u_{1p} \\
u_{4p} \\
u_{3p} \\
\end{array}
\right).
\end{equation}
The biorthogonality condition then reads:
\begin{equation}
\label{biorth}
\langle \vec{U}_p, \vec{V}_q \rangle =
N_p \delta_{pq},
\quad
N_p = \langle \vec{U}_p, \vec{V}_q \rangle.
\end{equation}
The norm $N_p$ can be calculated in closed form but these bulky expressions are not shown here.

\subsection{ Modes of vacuum waveguide }

It is useful to present TM and TE modes of vacuum waveguide in a similar form.
For $g=0$ and $\ve_{\bot} = \ve_{\parallel} = 1$ condition~\eqref{lambdasol} reduces to
\begin{equation}
\label{lambdasolvac}
J_0(as_e)J_1(as_o) = 0
\end{equation}
with immediate solution $s_{op} = j_{1p}/a$, $s_{ep} = j_{0p}/a$, where $J_0(j_{0p}) = 0$, $J_1(j_{1p}) = 0$.
In other words, we have two independent polarizations, TM and TE.
Let us present an eigenvector in the form:
\begin{equation}
\vec{U}_p =
\vec{U}_p^e + \vec{U}_p^h =
\left(
\begin{array}{c}
u_{1p}^e \\
u_{2p}^e \\
0 \\
0 \\
\end{array}
\right)
+
\left(
\begin{array}{c}
0 \\
0 \\
u_{3p}^h \\
u_{4p}^h \\
\end{array}
\right),
\end{equation}   
where superscript ``$e$'' means TM polarization while superscript ``$h$'' means TE polarization.
One obtains:
\begin{equation}
\vec{U}_p^e =
\left(
\begin{array}{c}
J_1\left( r\frac{j_{0p}}{a} \right) \\
\frac{\lambda_p^e}{k_0}J_1\left( r\frac{j_{0p}}{a} \right) \\
0 \\
0 \\
\end{array}
\right),
\,\,
\vec{U}_p^h 
=
\left(
\begin{array}{c}
0 \\
0 \\
J_1\left( r\frac{j_{1p}}{a} \right) \\
-\frac{\lambda_p^h}{k_0}J_1\left( r\frac{j_{1p}}{a} \right) \\
\end{array}
\right),
\end{equation}
where 
\begin{equation}
\label{lambdaeh}
\lambda_p^e = \sqrt{k_0^2 - \left( \frac{j_{0p}}{a} \right)^2},
\quad
\lambda_p^h = \sqrt{k_0^2 - \left( \frac{j_{1p}}{a} \right)^2}.
\end{equation}
Similarly to~\eqref{decomp},
\begin{equation}
\vec{\mathcal{E}}=
\vec{\mathcal{E}^e} + \vec{\mathcal{E}^h},
\end{equation}
where
\begin{equation}
\vec{\mathcal{E}^{e,h}}
=
\sum\limits_{p=-\infty}^{+\infty}
C_p^{e,h}\vec{U}_p^{e,h}(r)\exp\left[i\lambda_p^{e,h}z\right],
\end{equation}
and $C_p^{e,h}$ are some constants.

Formally, eigenvectors of the conjugate operator can be introduced by means of relation~\eqref{Vsol}:
\begin{equation}
\vec{V}_p^e =
\left(
\begin{array}{c}
\frac{\lambda_p^e}{k_0}J_1\left( r\frac{j_{0p}}{a} \right) \\
J_1\left( r\frac{j_{0p}}{a} \right) \\
0 \\
0 \\
\end{array}
\right),
\,\,
\vec{V}_p^h 
=
\left(
\begin{array}{c}
0 \\
0 \\
-\frac{\lambda_p^h}{k_0}J_1\left( r\frac{j_{1p}}{a} \right) \\
J_1\left( r\frac{j_{1p}}{a} \right) \\
\end{array}
\right),
\end{equation}
therefore the orthogonality conditions take the form:
\begin{equation}
\label{biorthvac}
\langle \vec{U}_p^e, \vec{V}_q^e \rangle =
N_p^e \delta_{pq},
\quad
\langle \vec{U}_p^h, \vec{V}_q^h \rangle =
N_p^h \delta_{pq},
\end{equation}
where
\begin{equation}
N_p^e = \frac{\lambda_p^e a^2}{k_0}J_1^2(j_{0p}),
\quad
N_p^h = -\frac{\lambda_p^h a^2}{k_0}J_0^2(j_{1p}).
\end{equation}

\section{ Boundary problem }

Now we return to the geometry shown in Fig.~\ref{fig:geom}.
Let us consider the case where a single mode of plasma-filled waveguide with number $l$ incidents the flat interface plasma -- vacuum located at $z=0$:
\begin{equation}
\label{ifield}
\vec{\mathcal{E}}^{(i)}
=
a^{(i)}
\vec{U}_l(r)\exp\left[i\lambda_l z \right].
\end{equation}
Reflected field in plasma is expressed as follows:
\begin{equation}
\label{rfield}
\vec{\mathcal{E}}^{(r)} = 
\sum\limits_{p=1}^{+\infty}
R_p\vec{U}_{-p}(r)\exp\left[ -i\lambda_p z \right],
\end{equation}
where $R_p$ are unknown reflection coefficients.
Transmitted field in vacuum is:
\begin{equation}
\label{tfield}
\vec{\mathcal{E}}^{(t)} = 
\sum\limits_{p=1}^{+\infty}
\left(
T_p^e\vec{U}_p^e(r)\exp\left[ i\lambda_p^e z \right]
+
T_p^h\vec{U}_p^h(r)\exp\left[ i\lambda_p^h z \right]
\right),
\end{equation}
where $T_p^{e,h}$ are unknown transmission coefficients (for two polarizations).
Continuity of tangential components of electric and magnetic fields at the interface $z=0$ results in the following relation:
\begin{equation}
\label{continuity}
a^{(i)}
\vec{U}_l(r)
+
\sum\limits_{p=1}^{+\infty}
R_p\vec{U}_{-p}(r) =
\sum\limits_{p=1}^{+\infty}
\left(
T_p^e\vec{U}_p^e(r) + T_p^h\vec{U}_p^h(r)
\right).
\end{equation}
Applying to~\eqref{continuity} 
$\langle \cdot, \vec{V}_{-q} \rangle$, 
$\langle \cdot, \vec{V}^e_{q} \rangle$
and 
$\langle \cdot, \vec{V}^h_{q} \rangle$, $q = 1, 2. \ldots$,
we arrive at the following relations:
\begin{equation}
\begin{aligned}
-R_q N_q &=
\sum\limits_{p=1}^{+\infty}
T_p^e \langle \vec{U}_p^e, \vec{V}_{-q} \rangle
+
\sum\limits_{p=1}^{+\infty}
T_p^h \langle \vec{U}_p^h, \vec{V}_{-q} \rangle, \\
T_q^e N_q^e &=
a^{(i)} \langle \vec{U}_l, \vec{V}^e_q \rangle
+
\sum\limits_{p=1}^{+\infty}
R_p \langle \vec{U}_{-p}, \vec{V}^e_q \rangle, \\
T_q^h N_q^h &=
a^{(i)} \langle \vec{U}_l, \vec{V}^h_q \rangle
+
\sum\limits_{p=1}^{+\infty}
R_p \langle \vec{U}_{-p}, \vec{V}^h_q \rangle.
\end{aligned}
\end{equation}
From these relations the following infinite linear system for $R_m$ can be obtained:
\begin{equation}
\sum\limits_{m=1}^{+\infty}
R_m
W_{mq}
+ R_q N_q = -a^{(i)}w_q,
\end{equation}
where
\begin{equation}
\begin{aligned}
W_{mq}= \sum\limits_{p=1}^{+\infty}\left[
\frac{ \langle \vec{U}_{-m}, \vec{V}^e_p \rangle \langle \vec{U}^e_p, \vec{V}_{-q} \rangle}{N_p^e}
\right. + \\
+
\left.
\frac{ \langle \vec{U}_{-m}, \vec{V}^h_p \rangle \langle \vec{U}^h_p, \vec{V}_{-q} \rangle}{N_p^h}
\right],
\end{aligned}
\end{equation}
\begin{equation}
\begin{aligned}
w_q= \sum\limits_{p=1}^{+\infty}\left[
\frac{ \langle \vec{U}_l, \vec{V}^e_p \rangle \langle \vec{U}^e_p, \vec{V}_{-q} \rangle}{N_p^e}
\right. + \\
+
\left.
\frac{ \langle \vec{U}_l, \vec{V}^h_p \rangle \langle \vec{U}^h_p, \vec{V}_{-q} \rangle}{N_p^h}
\right].
\end{aligned}
\end{equation}
The obtained system can be solved numerically.


%
%

\section{Conclusion}

We have constructed a rigorous solution for the problem of mode diffraction at the flat interface between magnetized plasma and vacuum inside a circular waveguide with perfectly conductive walls.
It should be underlined that we have considered the case of arbitrary external magnetic field $H_{ext}$.
First, modes of a plasma-filled have been obtained via eigenvectors of specific transverse operator: it operates with 4-vectors composed of transverse (with respect to waveguide axis $z$) components of EM field.
Biorthogonality condition for these eigenvectors has been obtained via eigenvectors of corresponding conjugated operator.
Solution of the boundary problem (reflection and transmission coefficient) has been presented as the solution of certain infinite linear system, the latter can be easily solved numerically

\section{Acknowledgments}
This work was supported by Russian Science Foundation (Grant No.~18-72-10137).
Author is grateful to S.~Baturin for fruitful discussions and to professor \fbox{V.V.~Novikov} for his outstanding lectures on EM propagation in plasma.

%

\end{document}